\def\fracm#1#2{\hbox{\large{${\frac{{#1}}{{#2}}}$}}}

\documentstyle[12pt]{article}

\def\@magscale#1{ scaled \magstep #1}

\catcode`@=11  
\def\un#1{\relax\ifmmode\@@underline#1\else
        $\@@underline{\hbox{#1}}$\relax\fi}
\catcode`@=12

\def\a{\alpha}
\def\b{\beta}

\def\d{\delta}
\def\e{\epsilon}

\def\g{\gamma}

\def\k{\kappa}
\def\l{\lambda}
\def\m{\mu}

\def\q{\theta}
\def\r{\rho}
\def\s{\sigma}

\def\D{\Delta}

\def\G{\Gamma}

\def\dslash{\not{\hbox{\kern-2pt $\partial$}}}
\def\Dslash{\not{\hbox{\kern-4pt $D$}}}
\def\pslash{\not{\hbox{\kern-2.3pt $p$}}}
 \newtoks\slashfraction
 \slashfraction={.13}
 \def\slash#1{\setbox0\hbox{$ #1 $}
 \setbox0\hbox to \the\slashfraction\wd0{\hss \box0}/\box0 }
 
\font\ro=cmsy10                          
\def\kcr{{\hbox{\ro \char'170}}}                
\def\ktl{{\hbox{\ro \char'170}}}        
\def\ktr{{\hbox{\ro \char'170}}}        
\def\kbl{{\hbox{\ro \char'170}}}        
\def\kbr{{\hbox{\ro \char'170}}}        

\def\plpl{\raise-2pt\hbox{$\raise3pt\hbox{$_+$}\hskip-6.67pt\raise0.0pt
\hbox{$^+$}\hskip 0.01pt$}}
\def\mimi{\raise-2pt\hbox{$\raise3pt\hbox{$_-$}\hskip-6.67pt\raise0.0pt
\hbox{$^-$}\hskip 0.01pt$}} 

\def\bo{{\raise.15ex\hbox{\large$\Box$}}}               
\def\pa{\partial}                                       
\def\TH{{\raise.2ex\hbox{$\displaystyle \bigodot$}\mskip-4.7mu \llap H
\;}}
\def\face{{\raise.2ex\hbox{$\displaystyle \bigodot$}\mskip-2.2mu \llap
{$\ddot
        \smile$}}}                                      

\def\sp#1{{}^{#1}}                              
\def\Tilde#1{\widetilde{#1}}                    
\def\Bar#1{\overline{#1}}                       
\def\leftrightarrowfill{$\mathsurround=0pt \mathord\leftarrow \mkern-6mu
        \cleaders\hbox{$\mkern-2mu \mathord- \mkern-2mu$}\hfill
        \mkern-6mu \mathord\rightarrow$}
\def\dvec#1{\vbox{\ialign{##\crcr
        \leftrightarrowfill\crcr\noalign{\kern-1pt\nointerlineskip}
        $\hfil\displaystyle{#1}\hfil$\crcr}}}           

\def\fracm#1#2{\hbox{\large{${\frac{{#1}}{{#2}}}$}}}
\def\frac#1#2{{\textstyle{#1\over\vphantom2\smash{\raise.20ex
        \hbox{$\scriptstyle{#2}$}}}}}                   
\def\sfrac#1#2{{\vphantom1\smash{\lower.5ex\hbox{\small$#1$}}\over
        \vphantom1\smash{\raise.4ex\hbox{\small$#2$}}}} 
\def\bfrac#1#2{{\vphantom1\smash{\lower.5ex\hbox{$#1$}}\over
        \vphantom1\smash{\raise.3ex\hbox{$#2$}}}}       
\def\afrac#1#2{{\vphantom1\smash{\lower.5ex\hbox{$#1$}}\over#2}}    

\newskip\humongous \humongous=0pt plus 1000pt minus 1000pt
\def\caja{\mathsurround=0pt}
\def\eqalign#1{\,\vcenter{\openup2\jot \caja
        \ialign{\strut \hfil$\displaystyle{##}$&$
        \displaystyle{{}##}$\hfil\crcr#1\crcr}}\,}
\newif\ifdtup

\def\ref#1{$\sp{#1)}$}

\parskip=\medskipamount                 
\thicklines                         

\def\oldheadpic{                                
        \setlength{\unitlength}{.4mm}
        \thinlines
        \par
        \begin{picture}(349,16)
        \put(325,16){\line(1,0){4}}
        \put(330,16){\line(1,0){4}}
        \put(340,16){\line(1,0){4}}
        \put(335,0){\line(1,0){4}}
        \put(340,0){\line(1,0){4}}
        \put(345,0){\line(1,0){4}}
        \put(329,0){\line(0,1){16}}
        \put(330,0){\line(0,1){16}}
        \put(339,0){\line(0,1){16}}
        \put(340,0){\line(0,1){16}}
        \put(344,0){\line(0,1){16}}
        \put(345,0){\line(0,1){16}}
        \put(329,16){\oval(8,32)[bl]}
        \put(330,16){\oval(8,32)[br]}
        \put(339,0){\oval(8,32)[tl]}
        \put(345,0){\oval(8,32)[tr]}
        \end{picture}
        \par
        \thicklines
        \vskip.2in}
\def\oldtitle#1#2#3#4{\oldheadpic\begin{center}\vglue.5in{\large\bf
#1}\\[.6in]
        {#2}\\[.1in] {\it Department of Physics and Astronomy}\\
        {\it University of Maryland, College Park, MD 20742}\\[.6in]
        Physics Publication \#{#3}\\ {#4}\\[1.5in] {\bf
ABSTRACT}\\[.1in]
        \end{center} \begin{quotation}}                 
\def\oldTitle#1#2#3#4#5#6#7{\oldheadpic\begin{center} \vglue .4in
        {\large\bf #1}\\[.4in]
        {#2}\\[.1in] {\it Department of Physics and Astronomy}\\
        {\it University of Maryland, College Park, MD 20742}\\[.1in]
        {#3}\\[.1in] {\it {#4}}\\ {\it {#5}}\\[.4in]
        Physics Publication \#{#6}\\ {#7}\\[.5in] {\bf ABSTRACT}\\[.1in]
        \end{center} \begin{quotation}}                 
\def\border{                                            
        \setlength{\unitlength}{1mm}
        \newcount\xco
        \newcount\yco
        \xco=-21
        \yco=12
        \begin{picture}(140,0)
        \put(\xco,\yco){$\ktl$}
        \advance\yco by-1
        {\loop
        \put(\xco,\yco){$\kcr$}
        \advance\yco by-2
        \ifnum\yco>-240
        \repeat
        \put(\xco,\yco){$\kbl$}}
        \xco=158
        \yco=12
        \put(\xco,\yco){$\ktr$}
        \advance\yco by-1
        {\loop
        \put(\xco,\yco){$\kcr$}
        \advance\yco by-2
        \ifnum\yco>-240
        \repeat
        \put(\xco,\yco){$\kbr$}}
        \put(-20,13){\tiny University of Maryland Elementary Particle
Physics University of Maryland Elementary Particle Physics University of
Maryland Elementary Particle Physics}
        \put(-20,-241.5){\tiny University of Maryland Elementary
Particle Physics University of Maryland Elementary Particle Physics
University of Maryland Elementary Particle Physics}
        \end{picture}
        \par\vskip-8mm}
\def\bordero{                                           
        \setlength{\unitlength}{1mm}
        \newcount\xco
        \newcount\yco
        \xco=-31
        \yco=12
        \begin{picture}(140,0)
        \put(\xco,\yco){$\ktl$}
        \advance\yco by-1
        {\loop
        \put(\xco,\yco){$\kclr}
        \advance\yco by-2
        \ifnum\yco>-240
        \repeat
        \put(\xco,\yco){$\kbl$}}
        \xco=151
        \yco=12
        \put(\xco,\yco){$\ktr$}
        \advance\yco by-1
        {\loop
        \put(\xco,\yco){$\kcr$}
        \advance\yco by-2
        \ifnum\yco>-240
        \repeat
        \put(\xco,\yco){$\kbr$}}
        \put(-20,12){\ooo
bacdefghidfghghdhededbihdgdfdfhhdheidhdhebaaahjhhdahba

hgdedge
   hgfdiehhgdigicba}
        \put(-20,-241.5){\ooo
ababaighefdbfghgeahgdfgafagihdidihiidhiagfedhadbfd

ecdcdfa
   gdcbhaddhbgfchbgfdacfediacbabab}
        \end{picture}
        \par\vskip-8mm}
\def\headpic{                                           
        \indent
        \setlength{\unitlength}{.4mm}
        \thinlines
        \par
        \begin{picture}(29,16)
        \put(165,16){\line(1,0){4}}
        \put(170,16){\line(1,0){4}}
        \put(180,16){\line(1,0){4}}
        \put(175,0){\line(1,0){4}}
        \put(180,0){\line(1,0){4}}
        \put(185,0){\line(1,0){4}}
        \put(169,0){\line(0,1){16}}
        \put(170,0){\line(0,1){16}}
        \put(179,0){\line(0,1){16}}
        \put(180,0){\line(0,1){16}}
        \put(184,0){\line(0,1){16}}
        \put(185,0){\line(0,1){16}}
        \put(169,16){\oval(8,32)[bl]}
        \put(170,16){\oval(8,32)[br]}
        \put(179,0){\oval(8,32)[tl]}
        \put(185,0){\oval(8,32)[tr]}
        \end{picture}
        \par\vskip-6.5mm
        \thicklines}
\def\title#1#2#3#4{\border\headpic {\hbox to\hsize{#4 \hfill UMDEPP
#3}}\par
        \begin{center} \vglue .5in {\large\bf #1}\\[.6in]
        {#2}\\[.1in] {\it Department of Physics and Astronomy}\\
        {\it University of Maryland, College Park, MD 20742}\\[1.5in]
        {\bf ABSTRACT}\\[.1in] \end{center} \begin{quotation}}  
\def\Title#1#2#3#4#5#6#7{\border\headpic
        {\hbox to\hsize{#7 \hfill UMDEPP #6}}\par
        \begin{center} \vglue .4in {\large\bf #1}\\[.4in]
        {#2}\\[.1in] {\it Department of Physics and Astronomy}\\
        {\it University of Maryland, College Park, MD 20742}\\[.1in]
        {#3}\\[.1in] {\it {#4}}\\ {\it {#5}}\\[.5in] {\bf
ABSTRACT}\\[.1in]
        \end{center} \begin{quotation}}                 
\def\endtitle{\end{quotation}\newpage}                  

\def\ad{{\kern0.5pt
                   \alpha \kern-5.05pt
\raise5.8pt\hbox{$\textstyle.$}\kern
0.5pt}}

\def\bd{{\kern0.5pt
                   \beta \kern-5.05pt
\raise5.8pt\hbox{$\textstyle.$}\kern
0.5pt}}

\def\qd{{\kern0.5pt
                   q \kern-5.05pt \raise5.8pt\hbox{$\textstyle.$}\kern
0.5pt}}
\def\Dot#1{{\kern0.5pt
                   {#1} \kern-5.05pt
\raise5.8pt\hbox{$\textstyle.$}\kern
0.5pt}}

\begin{document}

\def\gfrac#1#2{\frac {\scriptstyle{#1}}
        {\mbox{\raisebox{-.6ex}{$\scriptstyle{#2}$}}}}
\def\gg{{\hbox{\sc g}}}
\border\headpic {\hbox to\hsize{May 2002 \hfill
{UMDEPP 02-56}}}
\par
{\hbox to\hsize{$~$ \hfill
{CALT-68-2389}}}
\par
\setlength{\oddsidemargin}{0.3in}
\setlength{\evensidemargin}{-0.3in}
\begin{center}
\vglue .15in
{\large\bf Minimal Superspace Vector Fields for\\ 
5D Minimal Supersymmetry\footnote {Supported 
in part by National Science Foundation Grant PHY-98-02551}  }
\\[.5in]
S. James Gates, Jr.\footnote{gatess@wam.umd.edu}
\\
{\it Department of Physics, University of Maryland\\ 
College Park, MD 20742-4111 USA}\\ [0.08in]
and \\ [0.08in]
Lubna Rana\footnote{lubna@physics.umd.edu}
\\
{\it Physics \& Astronomy Department, Swarthmore College\\ 
500 College Ave, Swarthmore, PA 19081 USA}\\ [0.08in]

{\bf ABSTRACT}\\[.01in]
\end{center}
\begin{quotation}
{We investigate a minimal superspace description for 5D  
superconformal Killing vectors.  The vielbein appropriate 
for AdS symmetry is discussed within the confines of this minimal 
supergeometry.}  

\endtitle

\noindent
{\bf {(I.) Introduction}}  

With the proposal of the existence of the AdS/CFT correspondence
\cite{A}, the importance of understanding supersymmetry within 
the confines of a five dimensional spacetime became of major
importance.  Additionally, the introduction of five dimensional 
``semi-phenomenological'' models highlighted \cite{B,C} the 
importance of this class of models.  Most of these descriptions 
utilize a 5D superspace relying upon on Majorana-symplectic spinors 
to provide its fermionic coordinates.  If this were only description 
possible, then there could be no question of uniqueness.

In this letter, we wish to initiate study of an aspect of 5D models 
with simple supersymmetry that has been largely overlooked. To our
knowledge, all previous discussions of simple supersymmetry in 5D 
are based on the use of Majorana-symplectic spinors.  However, 
the use of Majorana-symplectic spinors is not dictated by any 
fundamental principle.  It is instead the legacy of the historical 
development of the field.   So we wish to begin to investigate the 
alternative choice where the 5D superspace does {\it {not}} rely 
upon Majorana-symplectic spinors to provide its fermionic coordinates.
\newline ${~~~~}$ \newline

\noindent
{\bf{(II.) The Spinor Metric and Gamma Matrices in Five Dimensions} }

In a five-dimensional Lorentzian spacetime, the minimal 
spinors are four component complex objects that we can 
denote by $\q^{\a}$.  We may choose a representation of 
the Dirac gamma matrices according to the conventions 
below.
$$ \eqalign{
(\gamma^{\un a})_{\a}{}^{\b} & \equiv ~ (~ \s^2 \otimes 
{\bf I}_2 \, , ~ i \s^3 \otimes {\bf I}_2 \, ,~ i \s^1 
\otimes \s^1 ,~ i \s^1 \otimes \s^2, ~ i \s^1 \otimes 
\s^3 ~) } \eqno(1) $$
These satisfy the equation
$$
\gamma^{\un a} \gamma^{\un b} ~+~  \gamma^{\un b} 
\gamma^{\un a} ~=~ - \, 2 \, \eta^{\un a \, \un b} \,{\bf I}_4
\eqno(2)$$
where the Minkowski space metric has the form,
$$
diag(\eta_{\un a \, \un b}) ~\equiv~ (~ -1, ~1,~1,~1,~1 ~) 
~~~.
\eqno(3)$$

One feature that we have traditionally used to simplify the
task of constructing spinor representations in various dimensions
is the introduction of a ``spinor metric.''  The usual gamma
matrices that appear in (1) and (2) may always be interpreted
as type (1,1) bispinors, i.e. one spinor index is up and one
is down. Matrix multiplication of such objects is always
well defined. Interpreting the gamma matrices as elements
of a Clifford algebra, together with the identity matrix and
all possible higher order anti-symmetrization of their
spacetime vector indices, it is always possible to find a
complete set $\G$ that provide a basis for all (1,1) bispinors.
Here we find
$$
\{ \G\}_{(1,1)} ~\equiv ~ \{~ ({\bf I})_{\a}{}^{\b}, ~ (\g^{\un 
a})_{\a}{}^{\b}, ~ (\s^{\un a \, \un b})_{\a}{}^{\b} ~ \} ~~~,
\eqno(4) $$
constitutes such a set. The set $\G_{(1,1)}$ is obviously a sum of
the $\{1\} \oplus \{5\} \oplus \{10\} $ irreducible representations
of the spacetime Lorentz group. The numbers of these matrices is
given by 1 + 5 + 10 = 16 and since our matrices are 4 $\times$ 4
we require 16 independent matrices for a complete basis. Next we 
introduce constant type (0,2) and (2,0) bispinors denoted by $\eta^{
\a \, \b}$ and $\eta_{\a  \, \b}$ that may be used to raise and lower 
spinor indices and such that $\eta_{\g \, \b} \eta^{\a \, \b} = 
\d_{\g}{}^{\a}$.  Now the defining properties of $\eta_{\a \, 
\g}$, the ``spinor metric'' and $\eta^{\a \, \g}$, the ``inverse 
spinor metric'' are that when they are used to form $\{ \G\}_{
(2,0)}$ and $\{ \G\}_{(0,2)}$, all elements of {\em {within}} each 
irreducible Lorentz representation are simultaneously members of
the {\em {same}} of representation of $S_2$ on the exchange of spinor 
indices.

A spinor metric may be introduced according to
$$
\eta_{\a \, \b} ~\equiv ~ (\s^3 \otimes \s^2)_{\a \, \b} ~~,~~
\eta^{\a \, \b} ~\equiv ~ - \, (\s^3 \otimes \s^2 )^{\a \, \b}
 ~~,~~ \eta_{\a \, \b} ~=~ - \, \eta_{\b \, \a} ~~,~~
\eta^{\a \, \b} ~=~ - \, \eta^{\b \, \a} ~~~.
\eqno(5)$$
This spinor metric may used to raise or lower one spinorial
index on the gamma matrices 
$$
(\gamma^{\un a})_{\a \, \b} ~\equiv~ (\gamma^{\un a})_{\a
}{}^{\g} \eta_{\g \, \b} ~~~,~~~ (\gamma^{\un a})^{\a \, 
\b} ~\equiv~  \eta^{\a \, \g} (\gamma^{\un a})_{\g}{}^{\b} ~~~,
\eqno(6)$$ 
whereupon we find,
$$
(\gamma^{\un a})_{\a \, \b} ~=~ - \, (\gamma^{\un a})_{\b \, 
\a} ~~~,~~~ (\gamma^{\un a})^{\a \, \b} ~=~ - \, (\gamma^{
\un a})^{\b \, \a}  ~~~.
\eqno(7)$$ 
The 5D spinor matrices whose commutator algebra is 
isomorphic to the Lorentz algebra is proportional to
$$
(\s^{\un a \, \un b})_{\a}{}^{\b} ~\equiv ~ i \, \fracm 12
\, (\, \gamma^{\un a}\gamma^{\un b } ~-~ \gamma^{\un b}
\gamma^{\un a} \, )_{\a}{}^{\b} ~~~.
\eqno(8)$$ 
The spinor metric may also used to raise or lower one spinorial
index on this quantity
$$
(\s^{\un a \, \un b})_{\a \, \b} ~\equiv~ (\s^{\un a
\, \un b})_{\a}{}^{\g} \eta_{\g \, \b} ~~~,~~~ (\s^{
\un a \, \un b})^{\a \, \b} ~\equiv~  \eta^{\a \, \g} 
(\s^{\un a \, \un b})_{\g}{}^{\b} ~~~,
\eqno(9)$$ 
whereupon we find,
$$
(\s^{\un a \, \un b})_{\a \, \b} ~=~  (\s^{\un a
\, \un b })_{\b \, \a} ~~~,~~~ (\s^{\un a \, \un b}
)^{\a \, \b} ~=~  (\s^{\un a \, \un b})^{\b \, \a}  ~~~.
\eqno(10) $$ 
So we have the following decomposition of $\G_{(2,0)}$
$$ \eqalign{ {~~~~~~~}
\G_{(2,0)} &=~ \G_{(2,0) \, S} ~ \oplus ~\G_{(2,0) \, A} ~~~, \cr
\G_{(2,0) \, S} &=~ \{~  (\s^{\un a \, \un b})_{\a \,\b} ~ \} 
~~~, ~~~ dim( \G_{(2,0) \, S} ) ~=~ \frac {4 \cdot 5}{1 \cdot 2} 
~=~ 10~~~, \cr
\G_{(2,0) \, A} &=~ \{~ \eta_{\a \, \b}, ~ (\g^{\un 
a})_{\a \, \b} ~ \} ~~~,  ~~~ dim( \G_{(2,0) \, A} )
 ~=~ \frac {4 \cdot 3}{1 \cdot 2} ~=~ 6 ~=~ 1 \, +\, 5 ~~~, 
}\eqno(11)$$
where to calculate the dimensionality of the symmetric and 
antisymmetric subspaces of $\G_{(2,0)}$ we simply have to take 
into account the symmetry or antisymmetry of the subspace and 
the dimensionality over which the spinor indices range.  Similar 
features are observed for $\G_{(0,2)}$.

Another useful object to introduce is a spinorial Levi-Civita
tensor $\e^{\a \, \b \, \g \, \d}$ with $\e^{1 \, 2 \, 3 \, 4
}\equiv + 1$. Again the spinor metric can lower indices
$$
\e_{\a \, \b \, \g \, \d} ~\equiv~ \e^{\k \, \l \, \phi 
\, \e} \eta_{\k \, \a} \eta_{\l \, \b} \eta_{\phi \, \g} 
\eta_{\e \, \d} ~~~,
\eqno(12) $$
so that
$$  
\e_{\a \, \b \, \g \, \d} ~\e^{\k \, \l \, \phi \, \e} 
~=~ \d_{[ \a |}{}^{\k} \, \d_{|\b | }{}^{\l} \, 
\d_{|\g | }{}^{\phi} \, \d_{ | \d ] }{}^{\e} ~~~. 
\eqno(13) $$
The spinorial $\e$-tensor also satisfies
$$ \eqalign{
\fracm 12 \, \e^{\a \, \b \, \g \, \d} \eta_{\g 
\, \d} &=~ \eta^{\a \, \b} ~~~,~~~ 
\fracm 12 \, \e^{\a \, \b \, \g \, \d} (\g_{\un 
a})_{\g \, \d} ~=~ - ~ (\g_{\un a})^{\a \, \b} 
 ~~~.
}\eqno(14) $$

The set of matrices given by 
$$
 ({\bf I}_4)_{\a}{}^{\b} ~~~,~~~  (\g^{\un a})_{\a}{}^{\b}
~~~,~~~ (\s^{\un a \, \un b })_{\a}{}^{\b} ~~~,
\eqno(15) $$
constitute a complete set of matrices over which type
(1,1) bi-spinors can be expanded. Similarly, the
following two sets are such bases for type (2,0)
and (0,2) bi-spinors, respectively.
$$
 ({\bf I}_4)_{\a \, \b} ~~~,~~~  (\g^{\un a})_{\a \, \b}
~~~,~~~ (\s^{\un a \, \un b })_{\a \, \b} ~~~,
\eqno(16) $$
$$
 ({\bf I}_4)^{\a \, \b} ~~~,~~~  (\g^{\un a})^{\a \, \b}
~~~,~~~ (\s^{\un a \, \un b \, })^{\a \, \b} ~~~.
\eqno(17) $$

There are also Fierz identities,
$$ \eqalign{ 
\e_{\a \, \b \, \g \, \d} &=\,  \fracm 12 \,\eta_{\a \, \b} 
\, \eta_{\g \, \d} \,+\, \fracm 12 \, (\gamma^{\un a})_{\a 
\, \b} (\gamma_{\un a} \, )_{\g \, \d} \,\,\,, \cr
\d_{[ \a}{}^{\g} \, \d_{ \b ]}{}^{\d} \, &=\, \fracm 12 \,
\eta_{\a \, \b} \, \eta^{\g \, \d} \,-\, \fracm 12 \, (
\gamma^{\un a})_{\a \, \b} (\gamma_{\un a} \, )^{\g 
\, \d} \,\,\,, \cr
\d_{( \a}{}^{\g} \, \d_{ \b )}{}^{\d} \, &=\, - \, \frac 
14 \, (\s^{\un a \, \un b })_{\a \, \b} (\s_{\un a \, \un 
b} \, )^{\g \, \d} \,\,\,, \cr
{~~~~~~~~}(\gamma^{\un a})_{\a}{}^{\, \g} (\gamma_{\un a} \,
)_{\b}{}^{\, \d} &=\,  - \fracm 14 \, [ \, 5 \eta_{\a \, \b} 
\, \eta^{\g \, \d} \,+\, 3 \, (\gamma^{\un a})_{\a \b} 
(\gamma_{\un a} \,)^{\g \d} \,-\, \frac 1{2} \, (\s^{\un a 
\, \un b })_{\a \, \b} (\s_{\un a \, \un b} \, )^{\g \, \d} 
\,]  ~\, .
} \eqno(18) $$

The representation of the 5D gamma-matrices that we have given also 
satisfies,
$$ \eqalign{
[ (\gamma^{\un a})_{\a}{}^{\b} ]^* &=~ + \, (C \, \gamma^{\un a}
\, C^{-1} )_{\a}{}^{\b} ~~~,~~~ C_{\a}{}^{\b} ~\equiv~ \s^1 
\otimes \s^2  ~~~, \cr 
~\to ~ C &=~ C^{-1} ~~~\&~~~ C_{\a}{}^{\g} 
\eta_{\g \, \b} ~=~ - \, C_{\b}{}^{\g} \eta_{\g \, 
\a} ~~~, \cr
[ (\s^{\un a \, \un b})_{\a}{}^{\b} ]^* &=~ - \, (C \, \s^{\un 
a \, \un b} \, C^{-1} )_{\a}{}^{\b} ~=~ -\, (C \, \s^{\un 
a \, \un b} \, C )_{\a}{}^{\b} ~~~.}
\eqno(19) $$

The multiplication table for the complete set of
4 $\times$ 4 matrices is given by
$$
\gamma^{\un a} \, \gamma^{\un b} ~=~  - \, \eta^{\un a \, 
\un b} \, {\bf I}_4 ~-~ i \, \s^{\un a \, \un b} 
~~~~, {~~~~~~~~~~~~~~~~~}{~~~~~~~~~~~~~~}
\eqno(20)$$
$$
\gamma^{\un a} \, \s^{\un b \, \un c} ~=~  - i \, \eta^{\un a
 \, [ \un b} \,  \gamma^{\un c ] } ~-~ \fracm 12  \, \e^{\un a 
\, \un b \, \un c \, \un d \, \un e} \, \s_{\un d \, \un e}
~~~~, {~~~~~~~~~~~~~~~~~~~}
\eqno(21)$$
$$
\s^{\un a \, \un b} \, \gamma^{\un c} ~=~ i \, 
\eta^{\un c \, [ \un a} \,  \gamma^{\un b ] } 
~-~ \fracm 12  \, \e^{\un a \, \un b \, 
\un c \, \un d \, \un e} \, \s_{\un d \, \un e}  
~~~~, {~~~~~~~}{~~~~~~~~~~~~~~}
\eqno(22)$$
$$ \eqalign{
\s^{\un a \, \un b} \, \s^{\un c \, \un d} &=~ 
\eta^{\un c \, \un a} \,  \eta^{\un b \, \un d} 
~-~ \eta^{\un d \, \un a} \, \eta^{\un b \, \un c} 
~-~  \e^{\un a \, \un b \, \un c \, \un d \, \un 
e} \, \g_{\un e} {~~~~~~~~~~~~~~~~~~~~} \cr
&{~~~} ~-~ i \, (\, \eta^{\un c \, [ \un a} 
\,  \s^{\un b ] \, \un d} ~-~ \eta^{\un d \, [ \un a}
\, \s^{\un b ]  \, \un c} \,) ~~~~.
}\eqno(23)$$

The spinorial supercovariant derivatives for the superspace
associated with these conventions can be defined by
$$
D_{\a} ~\equiv ~ \pa_{\a} ~+~ \fracm 12 \, (\g^{\un a})_{\a 
\, \b} C_{\g}{}^{\b} {\bar \q}^{\g} \, \pa_{\un a} ~~~,~~~
{\Bar D}{}_{\b} ~\equiv ~ C_{\b}{}^{\g} {\bar \pa}{}_{\g} ~-~ 
\fracm 12 \, (\g^{\un a})_{\b \, \g} {\q}^{\g} \, \pa_{\un a}
~~~. 
\eqno(24) $$
These satisfy the relations
$$ \eqalign{ {~~~~}
[~ D_{\a} ~ , ~ {\Bar D}{}_{\b} ~\} &=~  (\g^{\un a}
)_{\a \, \b}  \, \pa_{\un a} ~~~,~~~
[~ D_{\a} ~ , ~ {D}{}_{\b} ~\} ~=~ [~ {\Bar D}{}_{\a} ~ , 
~ {\Bar D}{}_{\b} ~\} ~=~ 0 ~~~, \cr
(D_{\a})^* &=~ - \, C_{\a}{}^{\b} \, {\Bar D}{}_{\b} ~~~,~~~ 
({\Bar D}{}_{\a})^* ~=~ - \, C_{\a}{}^{\b} \, {D}{}_{\b}~~~. 
}\eqno(25)$$
Note that the last equations imply
$$
((D_{\a})^*)^* ~=~ - \, D_{\a} ~~~,~~~ 
(({\Bar D}{}_{\a})^*)^* ~=~ - \, {\Bar D}{}_{\a} ~~~,
\eqno(26) $$
which is characteristic of a spacetime in which it is not possible
to define Majorana spinors.

At this point, it has long been the custom to append an SU(2) index 
to the spinor coordinates of the superspace ( $\q^{\a} \to \q^{\a 
\, i}$) and to further impose a ``Majorana'' condition on the enlarged 
spinorial coordinate,
$$
 (\q^{\a \, i})^* ~\equiv~ {\Bar \q}{}^{\a}_i ~=~ C_{i \, j} 
\q^{\a \, j}  ~~~,
\eqno(27) $$
which may be consistently imposed on the simplectic spinor
$\q^{\a \, i}$.  So the net effect of all of this is to first
double the number of independent spinorial coordinates ($\q^{
\a} \to \q^{\a \, i}$) and then cut this number in half (back 
to four independent spinorial coordinates) by the imposition 
of the condition in (27).  This naturally raises the question 
of what supergeometrical structures occur in the absence of
this process? \newline ${~~~~}$ \newline

\noindent
{\bf{(III.) Simple Poincar\' e Supervector Fields in 
Minimal 5D Superspace} }

The supersymmetry algebra associated with this minimal superspace can 
be read off from the results in (24) and (26).  The generators of the
minimal 5D superspace include a complex 4-component spinor $Q_{\a}$
(and ${\Bar Q}{}_{\a}$), the Lorentz 5-vector translation operator
$P_{\un a}$ and the second-rank Lorentz rotation generator $L_{\un 
a \, \un b}$.
$$ \eqalign{ {~~~}
P_{\un a} & \equiv ~ i \pa_{\un a} ~~~,~~~
L_{\un a \, \un b} ~\equiv~ i \, x_{[ \un a} \pa_{\un b ]} \,-\,
\frac 12 \, \q^{\a} (\s_{\un a \, \un b})_{\a}{}^{\b} \pa_{\b}
\,-\, \frac 12 \, {\bar \q}^{\a} (C \s_{\un a \, \un b} C)_{
\a}{}^{\b} {\bar \pa}_{\b} ~~~,~~~  \cr
Q_{\a} &\equiv ~ i \, [ ~\pa_{\a} ~-~  \fracm 12 \, (\g^{\un a})_{\a 
\, \b} C_{\g}{}^{\b} {\bar \q}^{\g} \, \pa_{\un a} ~] ~~~, ~~~
{\Bar Q}{}_{\b} ~\equiv ~ i\,  [~ C_{\b}{}^{\g} {\bar \pa}{}_{\g} ~+~ 
\fracm 12 \, (\g^{\un a})_{\b \, \g} \, {\q}^{\g} \, \pa_{\un a} ~] 
~~~. }
\eqno(29) $$ 
The only non-vanishing commutators are given by
$$ \eqalign{ {~~~~~~~~}
[~ Q_{\a} ~ , ~ {\Bar Q}{}_{\b} ~\} &=~ i \, (\g^{\un a})_{\a \, \b}
  \, P_{\un a} ~~~,~~~ 
[~ L_{\un a \, \un b} ~ , ~ Q_{\b} ~\} ~=~ \frac 12 \, (\s_{\un a 
\, \un b})_{\a}{}^{\b} Q_{\b} ~~~,~~~ \cr
[~ L_{\un a \, \un b} ~ , ~ {\Bar Q}{}_{\b} ~\} &=~ \frac 12 \, (
\s_{\un a \, \un b})_{\a}{}^{\b} {\Bar Q}{}_{\b}  ~~~,~~~
[~ L_{\un a \, \un b} ~ , ~ P_{\un c} ~\} ~=~ - i \, \eta_{\un c 
\, [ \un a} P_{\un b ] } ~~~,~~~ \cr 
[~ L_{\un a \, \un b} ~ , ~ L_{\un c \, \un d} ~\} &=~ - i \, 
\eta_{\un c \, [ \un a} L_{\un b ]  \, \un d} ~+~ i \, \eta_{\un 
d \, [ \un a} L_{\un b ]  \, \un c} ~~~,
}\eqno(30)$$
as expected.  So the super Poincar\' e group has a representation in 
terms of the supervector fields constructed from the derivatives with 
respect coordinates of the minimal 5D superspace. \newline ${~~~~}$ 
\newline

\noindent
{\bf{(IV.) Conformal Supervector Fields in Minimal 5D Superspace} }

To describe the superconformal algebra requires that we introduce
additional operators over and above those required to describe 
the super Poincar\' e algebra. In particular, we must introduce
$S_{\a}$, ${\Bar {S}}{}_{\a}$, $K_{\un a}$, $\D$, ${\cal J}^+$,
${\cal J}^-$ and ${\cal J}^0$ for the s-supersymmetry (and its 
conjugate), the special conformal boosts, dilatation generator and 
SU(2) charge operators, respectively.  The most interesting of the 
generators is the s-supersymmetry generator (and its conjugate) 
defined by, 
$$ \eqalign{ {~~~~~~}
{S}_{\a} &\equiv~  [~ - i x^{\un a} (\g_{\un a})_{\a}{}^{\b} \, 
{\Bar Q}{}_{\b}  \,+\, C_{\r}{}^{\d} {\bar \q}{}_{\d} \, (\g^{\un 
a} )^{\r}{}_{\s} \, \q^{\s} \, \q_{\a} \pa_{\un a} \,+\, \fracm 14 
(\g^{\un a} C )_{\a}{}^{\d}{\bar \q}{}_{\d} \, \q^{\r}  \q_{\r} 
\pa_{\un a}\cr 
&{~~~~~~} ~+~ \fracm 34 C_{\r}{}^{\d} {\bar \q}{}_{\d} \, 
\q^{\r} (\g^{\un a} )_{\a \s} \q^{\s} \pa_{\un  a}\,+\, \q_{
\a}{\bar \q}{}^{\r} {\bar \pa}{}_{\r} \,-\, \fracm 12 C_{\d}
{}^{\s}{\bar \q}{}_{\s} \, \q^{\d} C_{\a}{}^{\g}{\bar \pa}{
}_{\g} \cr
&{~~~~~~\,} \,-\, 2 C_{\a}{}^{\d} {\bar \q}{}_{\d} \, \q^{\e} 
C_{\e}{}^{\g}  {\bar \pa}{}_{\g} \,-\, \q_{\a} \, \q^{\r} \pa_{\r} 
\,+\, \fracm 12 \q^{\r} \, \q_{\r} \pa_{\a} ~] ~~~,~~~ } 
\eqno(31) $$
$$
\eqalign{ {~~~~~} 
{\Bar {S}}{}_{\a} \,&\equiv~  s_1 [~ - i x^{\un a} (\g_{\un a}
)_{\a}{}^{\b} \, {Q}{}_{\b} \,+\, \fracm 14   (\g^{\un 
a} )_{\a}{}^{\d} {\q}{}_{\d} \, {\bar \q}{}^{\r}  {\bar \q}{}_{\r} 
\pa_{\un a} \,+\,  {\q}{}_{\r} \, (  \g^{\un a}  C )^{\r}{}_{\s} 
\, {\bar \q}{}^{\s} \,  C_{\a}{}^{\g}{\bar \q}{}_{\g} \pa_{\un a}\cr 
&{~~~~~~} ~+~ \fracm 34  C_{\r}{}^{\d} {\q}{}_{\d} \, 
{\bar \q}{}^{\r} (\g^{\un a} C)_{\a \s} {\bar \q}{}^{\s} \pa_{\un 
a}\,-\,  C_{\a}{}^{\g}
{\bar \q}{}_{\g}{\q}{}^{\r} {\pa}{}_{\r} \,+\, \fracm 12  C_{\d}
{}^{\s}{\bar \q}{}_{\s} \, \q^{\d} {\pa}{}_{\a} \cr
&{~~~~~~\,} \,+\, 2  {\q}{}_{\a} \, {\bar \q}{}^{\e} C_{\e}{
}^{\g}  {\pa}{}_{\g} \,+\,  C_{\a}{}^{\g}{\bar \q}{}_{\g} \, 
{\bar \q}{}^{\r} {\bar \pa}{}_{\r} \,-\, \fracm 12 {\bar 
\q}{}^{\r} \, {\bar \q}{}_{\r} C_{\a}{}^{\g}{\bar \pa}{}_{\g} ~] 
~~~,~~~ } \eqno(32)$$
The coefficient $s_1 = \pm 1$ for the present.  We also require
additional bosonic generators\footnote{The terms in the
ellipsis below can explicitly be determined from the first 
result in (34).}.
$$
\eqalign{ {~~~~~} 
K_{\un a} \,& \equiv ~ i [ \, x^2 \, \pa_{\un a} ~-~ 2 \, x_{\un a}
x^{\un b} \pa_{\un b} ~+~.... \,] ~~~,~~~ \cr
\D \,&\equiv~ i \, [\,  x^{\un a} \pa_{\un a} ~+~ \frac 12 \, \q^{\a}
\pa_{\a} ~+~ \frac 12 \, {\bar \q}^{\a} {\bar \pa}_{\a} \,]
~~~,~~~ \cr
{\cal J}^0 \,&\equiv~  \fracm 12 \, [\, \q^{\a} \pa_{\a} ~-~ {\bar 
\q}^{\a} {\bar \pa}_{\a} \,] ~~~,~~~ 
{\cal J}^+ \,~\equiv~  \fracm 1{\sqrt 2} \, [\, {\bar \q}{}^{\a} 
C_{\a}{}^{\b} {\pa}_{\b} \,] ~~~,~~~ \cr
{\cal J}^- \,&\equiv~ \fracm 1{\sqrt 2} \, [\, \q^{\a} C_{\a}{}^{\b} 
{\bar \pa}{}_{\b} \,] 
 ~~~.}
\eqno(33) $$

The full commutator algebra with the choice $s_1 = -1$ is given below:
$$ \eqalign{ {~~~~~~}
[~ S_{\a} ~ , ~ {\Bar S}{}_{\b} ~\} \, &=~ +i (\g^{\un a})_{\a 
\, \b}  \, K_{\un a} ~~~\,~~,~~~~~ [~ L_{\un a \, \un b} ~ , ~ K_{\un 
c} ~\} \, ~=~ - i \, \eta_{\un c \, [\un a} K_{\un b ] } ~~~~, \cr
[~ L_{\un a \, \un b} ~ , ~ S_{\a} ~\} \, &=~ \frac 12 \, (\s_{
\un a \, \un b})_{\a}{}^{\b} S_{\b} ~~~\,~,~~~~~ 
[~ L_{\un a \, \un b} ~ , ~ {\Bar S}{}_{\a} ~\} \, ~=~ \frac 12 
\, ( \s_{ \un a \, \un b})_{\a}{}^{\b} {\Bar S}{}_{\b} ~~~,~~~ \cr
[~ S_{\a} ~ , ~ P_{\un a} ~\} \, &=~ -  (\g_{\un a})_{\a}{}^{\g}
{\Bar Q}{}_{\g} ~~~, ~~~~~ [~ {\Bar S}{}_{\a} ~ , ~ P_{\un a} ~\} \, 
~=~ + (\g_{\un a})_{\a}{}^{\g} {Q}{}_{\g} ~~~, \cr
[~ Q_{\a} ~ , ~ K_{\un a} ~\} \, &=~ + (\g_{\un a})_{\a
}{}^{\g} {\Bar S}{}_{\g}  ~~~,~~~ [~ {\Bar Q}{}_{\a} ~ , ~ 
K_{\un a} ~\} \, ~=~ - (\g_{\un a})_{\a}{}^{\g} {S}{}_{\g}   ~~~~\,,
\cr
[~ Q_{\a} ~ , ~ S_{\b} ~\} \, &=~ - i \fracm 12 (\s^{\un a 
\, \un b})_{\a \, \b} \, L_{\un a \, \un b} ~+~  \eta_{\a \,
\b} \, [ ~\D ~+~  3 \, {\cal J}^0 ~ ] ~~~, {~~~~~~~~~~~~} \cr
[~ {\Bar Q}{}_{\a} ~ , ~ {\Bar S}{}_{\b} ~\} \, &=~ - i   \fracm 12 
(\s^{\un a \, \un b})_{\a \, \b} \, L_{\un a \, \un b} ~-~ s_1  \eta_{
\a \, \b} \, [ ~\D ~+~  3 \, {\cal J}^0 ~ ] ~~~~~~~, {~~~~~~~} \cr
[~ \D ~ , ~ Q_{\a} ~\} \, &=~ - i \, \fracm 12  \, Q_{\a} ~~~~~~, ~~~~~ 
[~ \D ~ , ~ S_{\a} ~\} \, ~=~ i \fracm 12 \, S_{\a} ~~~~~~\,~, \cr
[~ \D ~ , ~ {\Bar Q}{}_{\a} ~\} \, &=~ - i \fracm 12 \, {\Bar Q}_{\a} 
~~~~~~, ~~~~~ [~ \D ~ , ~ {\Bar S}{}_{\a} ~\} \, ~=~ i \fracm 12 
\, {\Bar S}{}_{\a} ~~~~~~~, \cr
[~ {\cal J}^0 ~ , ~ Q_{\a} ~\} \, &=~  - \fracm 12 \, Q_{\a} ~~~~~~, 
~~~~~ [~ {\cal J}^0 ~ , ~ S_{\a} ~\} \, ~=~  \fracm 12 \,  S_{\a} 
~~~~\,~~~, \cr
[~ {\cal J}^0 ~ , ~ {\Bar Q}{}_{\a} ~\} \, &=~  \fracm 12  \, {\Bar Q}
{}_{\a} ~~~~~~~~~~, ~~~~~ [~ {\cal J}^0 ~ , ~ {\Bar S}{}_{\a} ~\} \, ~=~  
- \fracm 12 {\Bar S}{}_{\a} ~~~~~~~, \cr
[~ {\cal J}^+ ~ , ~ Q_{\a} ~\} \, &=~  0 ~~~\,~~~~~~~~~~ ~,~~~~~~
[~ {\cal J}^+ ~ , ~ S_{\a} ~\} \, ~=~ +  \fracm 1{\sqrt 2} 
{\Bar S}_{\a} ~~~~\,~~~, \cr
[~ {\cal J}^+ ~ , ~ {\Bar Q}{}_{\a} ~\} \, &=~  - \fracm1{\sqrt 2} \, 
{Q}{}_{\a} ~~~~~~, ~~~~~ [~ {\cal J}^+ ~ , ~ {\Bar S}{}_{\a} ~\} \, ~=~  
0 ~~~~~~~, \cr
[~ {\cal J}^- ~ , ~ Q_{\a} ~\} \, &=~ - \fracm1{\sqrt 2} \, {\Bar 
Q}{}_{\a} ~~~~~~, ~~~~~ [~ {\cal J}^- ~ , ~ S_{\a} ~\} \, ~=~  
0 ~~~~\,~~~, \cr
[~ {\cal J}^- ~ , ~ {\Bar Q}{}_{\a} ~\} \, &=~  0
 ~~~~~~~~~~~\,~~~~~, ~~~~~ [~ {\cal J}^- ~ , ~ {\Bar S}{}_{\a} ~\} \, ~=~  
+  \fracm 1{\sqrt 2} {S}{}_{\a} ~~~~~~~, \cr
[~ \D ~ , ~ P_{\un a} ~\} \, &=~ -i \, P_{\un a} ~~~\,~~~~~~, ~\,~~~~
[~ \D ~ , ~ K_{\un a} ~\} \, ~=~ i \, K_{\un a} ~~~~\,~~~, \cr
[~ P_{\un a} ~ , ~ K_{\un b} ~\} \, &=~  i 2 \, L_{\un a 
\, \un b} ~-~ i 2  \eta_{\un a \, \un b} \, [ ~  \D ~+~  3 
{\cal J}^0 ~ ] ~~~, {~~~~~~~~~~~~} \cr
[~ {\Bar Q}{}_{\a} ~ , ~ {S}{}_{\b} ~\} \, &=~ - i 3 \sqrt 2 \,\eta_{
\a \b} \, {\cal J}^- ~~~,~~~[~ {Q}{}_{\a} ~ , ~ {\Bar S}{}_{\b} ~\} \, 
~=~  - i  3 \sqrt 2 \,\eta_{\a \b} \, {\cal J}^+ ~~~, \cr
[~ {\cal J}^+ ~ , ~ {\cal J}^0 ~\} &=~ {\cal J}^- ~~~,~~~
[~ {\cal J}^0 ~ , ~ {\cal J}^- ~\} ~=~ {\cal J}^+ ~~~.
[~ {\cal J}^- ~ , ~ {\cal J}^+ ~\} \, ~=~ {\cal J}^0  ~~~.}
\eqno(34) $$
The major feature that emerges from the construction of the
supervector fields that generate the superconformal group is
the unavoidable appearance of a triplet of SU(2) generators
(${\cal J}^+$, ${\cal J}^0$, ${\cal J}^-$).   
\newline ${~~~~}$ \newline

\noindent
{\bf{(V.) Simple AdS Supervector Fields in Minimal 5D Superspace} }

Once the supervectors fields for a conformal group are known for a 
superspace, there follows as an immediate consequence the possibility
of realizing an anti-de-Sitter (AdS) supersymmetry.  The key point is 
that the AdS supersymmetry generators (${\cal Q}$) and AdS translation 
operators (${\cal P}$) can be defined in terms of operators for the
conformal group by the equations
$$
{\cal Q}_{\a} ~=~ Q_{\a} ~+~ \l \, {\Bar S}{}_{\a} ~~~,~~~ {\Bar 
{\cal Q}}{}_{\a} ~=~ {\Bar Q}{}_{\a} ~-~ s_1 {\bar \l} \,  {S}{}_{\a} 
~~~,~~~ {\cal P}_{\un a} 	~=~ P_{\un a}
~+~ |\l|^2 \, K_{\un a} ~~~.
\eqno(35) $$
All of these features were discussed many years ago in {\it {Superspace}}.  
The non-vanishing terms in the 5D AdS supersymmetry algebra take the 
form\footnote{The parameter $s_1$ may conveniently be set equal to
minus one.} 
$$ \eqalign{ {~~~~}
[~ {\cal Q}_{\a} ~ , ~ {\Bar {\cal Q}}{}_{\b} ~\} &=~ i \, (\g^{\un a}
)_{\a \, \b}  \, {\cal P}{}_{\un a} ~+~ i s_1 \, (\, \l \,+\, {\bar \l}
\, ) \,
\, (\s^{\un a \, \un b})_{\a \, \b} \, L_{\un a \, \un b} \cr
& ~~~~~+~ s_1 \,  ( \l \,-\, {\bar \l})\, {\eta}_{\a \,
\b} \, (\, \D  ~+~ 3 {\cal J}^0 \,)
~~~,~~~ \cr
[~ {\cal P}{}_{\un a}  ~ , ~ {\cal P}{}_{\un b} ~\} &=~ i 4 \, | \l 
|^2 \, L_{\un a \, \un b}
~~~. }
\eqno(36) $$
The second of these has the characteristic form of the algebra of
translation operators in a space of constant negative curvature.
One of the uses of this information is that it allows the construction
of the rigid AdS limit of five dimensional supergravity in minimal 
superspace.  This is done first by noting that the supergravity 
vielbein superfield can be directly defined in terms of the AdS 
supervector fields defined immediately above. 

The first step in this construction is to introduce a AdS 
background-fixed vielbein superfield ${\Tilde {\rm E}}
{}_{\un A}{}^{\un M}(\q, \, {\bar \q} , \, x) $.  We next observe 
that there exist a canonical relation between the supervector fields 
above and the AdS background-fixed vielbein superfield. This 
relation is,
$$ \eqalign{
{\cal Q}_{\a} &\equiv~ i \,{\Tilde {\rm E}}{}_{\a}{}^{\un M} (- \q, 
\, - {\bar \q} , \, x)  \, D_{\un M}  ~~~,~~~  {\cal P}_{\un a} 
~\equiv~ i \, {\Tilde {\rm E}}{}_{\un a}{}^{\un M} (- \q, \, - {\bar \q} 
, \, x)
D_{\un M} ~~~.
} \eqno(37) $$
Upon solving both of the above equation for ${\Tilde {\rm E}}{}_{\un 
A}{}^{\un M}(\q, \, {\bar \q} , \, x)$, one obtains an explicit expression 
for the vielbein superfield.  Since the superconformal vector fields are 
$\l$-dependent, the AdS background-fixed vielbein superfield ${\Tilde {\rm 
E}}{}_{\un A}{}^{\un M}$ is also $\l$-dependent.  The general superspace 
measure (``D-term'') measure for the five dimensional superspace of 
minimal supersymmetry is then
$$
\int d \m_{AdS} ~=~ \int d^5 x \, d^4 \q \, d^4 {\bar \q} 
~ [ sdet ({\Tilde {\rm E}}{}_{\un A} {}^{\un M} )]^{-1} ~~~.
\eqno(38) $$
There remains the separate problem of how to construct the AdS
superspace chiral measure.  This is a problem that will have to
be addressed in a future work.
\newline ${~~~~}$ \newline

\noindent
{\bf{(VI.) Discussion and Future Prospectives} }

As we have seen, there appear no major impediments to a hitherto 
unconventional description of simple supersymmetry in a five
dimensional superspace.  The formalism that we have established
is based on the fact that complex spinors 4-component spinors
provide a perfectly adequate basis for describing the fermionic
components of a 5D superspace.  The super Poincare\' e group 
associated with this construction is consistent with either a
U(1) or SU(2) holonomy.  The superconformal group associated
with this construction requires a triplet of SU(2) generators to close.

One of our purposes to studying the realization of superconformal
vector fields in the present context was in preparation for an
investigation on how superconformal symmetry is {\em {geometrically}}
{\em {realizable}} (${\cal {GR}}$) in all superspaces.  Between the 
experience gained in the present investigation and that of some 
previous works, one is led to expect that superconformal Killing
vector {\em {can}} be realized in all dimensions!  In terms of
supergeometry, there is every reason to expect that 10D and 11D
generalizations of the results in (29), (32) and (33) should
exist.

This statement is at variance with long held beliefs about this topic.  
There are ``no-go'' theorems which purport to limit the spacetime 
dimensions in which superconformal symmetry can appear.  In fact, an 
examination of the purported proofs reveals that {\em {none}} of them 
are based on the vector bundles naturally associated with the differential 
geometry of Salam-Strathdee superspace.  As such these ``proofs'' have 
no {\em {a}} {\em {priori}} validity with regard to placing 
restrictions on the representations carried by superfields.  As is 
often the case for no-go theorems in mathematical physics, their 
restrictions can often be avoided by relaxing one or more of the 
assumptions that go into their construction.  The caveat that allows 
for the existence of geometrically realizable superconformal symmetries 
for all superspaces is {\em {likely}} the existence of new charges 
in the algebra that are multi-spinor representations of the Lorentz 
sub-algebra.  

This solution has been hinted at by much of the recent literature on
conformal symmetry especially in 11D supersymmetric models.  The
extra 2-form and 5-form charges to which appropriate branes are 
expected to couple play this role.  In the model-independent 
description of centerless super-Virasoro algebras \cite{D}, in
particular, new charges (called the ``$U$'' and ``$R$'' charges) 
must occur.  Relating these 1D charges to the 11D 2-brane and 5-brane 
charges seems a worthy task.  Upon reduction to 1D, all the generators 
of the higher D ${\cal {GR}}$ superconformal algebra may be expected 
to become the complete set of generators for the 1D $N$-extended super
${\cal {GR}}$ Virasoro algebras.

${~~~}$ \newline
${~~~~~~~~~}$``{\it {The absurd has meaning only in so far as it
is not agreed to}}'' -- Albert Camus.

$${~~~}$$
\noindent


\begin{thebibliography}{66}

\bibitem{A}J.~Maldacena, Adv.~Theor.~Math.~Phys. {\bf 2} (1998) 231.

\bibitem{B}P.~Ho\v rava and E.~Witten, Nucl.~Phys.~{\bf {B460}} (1998) 506.

\bibitem{C}E.A.Mirabelli and M.E.Peskin, Phys.~Rev.~{\bf {D58}}
(1998) 065002.

\bibitem{D}S.\ J.\ Gates, Jr.\ and L.\ Rana, Phys.~Lett.~{\bf {438B}} 
(1998) 80 (hep-th/9806038); C.\ Curto, S.\ J.\ Gates, Jr.\  and V.\ G.\ J.\ Rodgers,
Phys.~Lett.~ {\bf {450B}} (2000) 337 (hep-th/0002010); S.\ J.\ Gates, 
Jr.\  and V.\ G.\ J.\ Rodgers, Phys.\ Lett.\ B512 (2001) 189 
(hep-th/0105161); A.\ Boviea,  S.\ J.\ Gates, Jr., D.\ M.\ Kimberly, B.\
A.\ Larson and V.\ G.\ J.\ Rodgers, Phys.~Lett.~{\bf {529B}} (2002) 222
(hep-th/0201094).


\end{thebibliography}
\end{document}